\documentclass{article}
\PassOptionsToPackage{numbers, compress}{natbib}
\usepackage[final]{nips_2018}
\usepackage[utf8]{inputenc} 
\usepackage[T1]{fontenc}    
\usepackage{hyperref}       
\usepackage{url}            
\usepackage{booktabs}       
\usepackage{amsfonts}       
\usepackage{nicefrac}       
\usepackage{microtype}      
\usepackage{amsmath}
\usepackage{xcolor}
\usepackage{graphicx}
\usepackage{graphics}
\usepackage{amsmath}
\usepackage{float}
\title{Dual Objective Approach Using A Convolutional Neural Network for Magnetic Resonance Elastography}

\author{
  Ligin M. Solamen\\
  Thayer School of Engineering\\
  Dartmouth College\\
  Hanover, NH 03755\\
  \texttt{ligin.m.solamen.th@dartmouth.edu} \\
  \And
    Yipeng Shi \\
    Department of Physics \\
    Dartmouth College\\
    Hanover, NH 03755\\
    \texttt{yipeng.shi.gr@dartmouth.edu} \\
   \And
   Justice Amoh Jr.  \\
   Thayer School of Engineeing \\
   Dartmouth College \\
   \texttt{justice.amoh.jr.th@dartmouth.edu} \\
}

\begin{document}

\maketitle

\begin{abstract}
  Traditionally, nonlinear inversion, direct inversion, or wave estimation methods have been used for reconstructing images from MRE displacement data. In this work, we propose a convolutional neural network architecture that can map MRE displacement data directly into elastograms, circumventing the costly and computationally intensive classical approaches. In addition to the mean squared error reconstruction objective, we also introduce a secondary loss inspired by the MRE mechanical models for training the neural network. Our network is demonstrated to be effective for generating MRE images that compare well with equivalents from the nonlinear inversion method.   
\end{abstract}

\section*{Background}
Palpation, the technique of evaluating tissue in the human body through touch, is widely used in medicine. It can help assess tenderness, swelling, and stiffness changes. However, palpation is limited by its qualitative measurement, restricted scope to only examine superficial organs, and the inability to precisely compare results between patients. Quantitatively, palpation is measuring the elastic modulus, a mechanical property that assess the resistance of deformation when stress is applied. Elastography, a method to quantitatively measure the mechanical properties of tissue, was proposed as a way to overcome the challenges of palpation and provide clinically relevant mechanical property information.  \citet{ophir1991elastography} presented the technique and the first results of imaging the elastic modulus of soft tissue using ultrasound termed ultrasound elastography. \cite{ophir1996elastography,ophir1999elastography} Elastography was expanded to magnetic resonance given its advantage of high resolution imaging. \cite{muthupillai1995magnetic,muthupillai1996magnetic} Magnetic resonance elastography (MRE) allowed for the non-invasive study of nearly all tissue elastic properties.  

Gathering tissue deformation and stiffness information can be useful in the diagnosis and treatment planning of various cancers. For example, thyroid cancer has a clinical presentation of firm consistency of the thyroid gland. \cite{mangione2000physical} Similarly, breast cancer initiation and progression is marked by structural changes in the extracellular matrix resulting in changes of stiffness and other mechanical properties of the soft glandular tissue. \cite{plodinec2012nanomechanical} The development of MRE enabled the exploration of subtle mechanical property changes in tissue. MRE in liver diseases have been successful; staging hepatic fibrosis using MRE is clinically used as an alternative to biopsy. \cite{venkatesh2014magnetic}

The development of MRE enabled the exploration of mechanical property changes in neurological disorders. Multiple sclerosis, a demyelinating disease, decreases the tissue stiffness caused by the alterations in the brain parenchyma. \cite{wuerfel2010mr, streitberger2012brain} Patients with Alzheimer's Disease, a progressive disease that causes decline in cognitive function, were found to have a decrease in brain tissue stiffness compared to healthy controls. \cite{murphy2011decreased}  MRE has also shown promising results in the assessment of meningiomas, a fibrous intracranial tumor, and outperformed conventional preoperative assessment imaging in predicting tumor stiffness and intratumoral consistency. \cite{murphy2013preoperative,hughes2015higher} MRE has a clear potential to add value to the identification and management of patients with neurological disorders.

\section*{Introduction}
MRE is an imaging technique that involves a three step process. First, mechanical actuators introduce harmonic motion (on the order of 10-1000$\mu$m and at a single frequency between 25-100Hz for \textit{in vivo} imaging) to generate shear waves that propagate into the tissue of interest. Second, MRI measures the resulting displacement with specialized pulse sequences that incorporate motion encoding gradients. \cite{weaver2001magnetic} In the brain, external motion actuators can be eliminated and motion sensitive MR pulse sequences can be tuned to capture the intrinsic motion of the brain caused blood pressure variations during the cardiac cycle. \cite{weaver2012brain} Finally, measured displacements are used to estimate mechanical property images, called elastograms, primarily stiffness represented by the complex shear modulus, $\mu$.  

Estimating accurate and relevant mechanical properties, an inverse problem, is a large obstacle in MRE and an active area of research.  \cite{van2000elasticity, van2001three, perrinez2009modeling, mariappan2010magnetic, sack2013structure,guo2013towards} Inversion algorithms vary by their underlying mechanical assumptions, actuation frequency, complexity of numerical estimation, and computational time. Applications of machine learning in the field of MRE is appealing because it removes many constraints of the inversion algorithm such as an initial guess of estimated properties, spatial filtering, partial volume effects, actuation frequency limitations on mechanical models, and model data mismatch for different tissues. \cite{mcgarry2015suitability} Recently, \citet{murphy2018artificial} presented the first work in the application of deep learning techniques, artificial neural networks, in the field of MRE. In this paper, we further that effort by investigating convolutional neural networks (CNNs) as a more suitable architecture for mapping motion data to stiffness elastograms. Using simulated MRE data, our proposed CNN model is shown to generate elastograms that compare well with those from non-linear inversion methods. 




\section*{MRE Dataset}
For training our neural network, MRE simulation data is generated using a poroelastic  mechanical model (PE) described by the following equations:  \cite{perrinez2009modeling,perrinez2010contrast,pattison2014spatially}

\begin{subequations} 
\centering
\begin{align}
\nabla\cdot(\mu_p(\nabla\overrightarrow{u}+\nabla\overrightarrow{u}^T))+\nabla(\lambda\nabla\cdot\overrightarrow{u})-(1-\beta)\nabla p &=-\omega^2(\rho-\beta\rho_f)\overrightarrow{u}     \label{poroeq_a}\\   
\nabla\cdot(\beta\nabla p) + \omega^2\rho_f\nabla\cdot((1-\beta)\overrightarrow{u}) &= 0   \label{poroeq_b}  \\   
\beta&=\frac{\omega\phi^2_p\rho_f\kappa}{i\phi^2_p+\kappa\omega(\rho_a+\phi_p\rho_f)}   
\label{poroeq_c}   
\end{align}
\end{subequations}

\noindent where $\mu$ is the shear modulus for the porous matrix,
$\overrightarrow{u}$ is the complex-valued 3D displacement vector, $\lambda$ is the first Lam\'{e} constant, $p$ is the pore pressure, $\omega$ is the actuation frequency, $\rho$ is the solid density, $\rho_f$ is the fluid density, $\rho_a$ is the apparent mass density, $\phi_p$ is the porosity, and $\kappa$ is the hydraulic conductivity. Hydraulic conductivity describes the ease of fluid flow through the porous matrix and porosity is the ratio of pore space to total volume.  \cite{pattison2014spatially} The assumed constants for the  model were: $\kappa$=1e$^-8$ $m^3s/kg$, $\phi_p$=0.2, $\rho$=1020 $kg/m^3$, $\rho_f$=1000 $kg/m^3$,  $\rho_a$=150 $kg/m^3$. \cite{perrinez2009modeling}.  A PE model describes the behavior of a biphasic media that is comprised of both a solid and fluid phases that interact in a coupling phenomena where applied stress causes a change in mass or fluid pressure and change in fluid pressure or mass results in volume change of the solid matrix. \cite{perrinez2009modeling,perrinez2010contrast} 

The poroelastic forward solution (estimating $\overrightarrow{u}$  given spatially varying $\mu$ and appropriate pressure boundary conditions) produced complex valued 3D displacements for a 1 mm harmonic actuation along the first direction. The simulated data comprises of eight basic 3D configurations or objects (eg. brain, cylinders and prisms), with each resolving to at least 14 2D slices. Additionally, simulations are performed at multiple actuation frequencies, $\omega$, in the range of 1-200Hz. The overall dataset eventually consists of 1,587 images in total. The complex-valued displacement images, $\overrightarrow{u}$, was transformed into absolute displacements. 

\begin{figure}
\centering
  \includegraphics[width=1\textwidth]{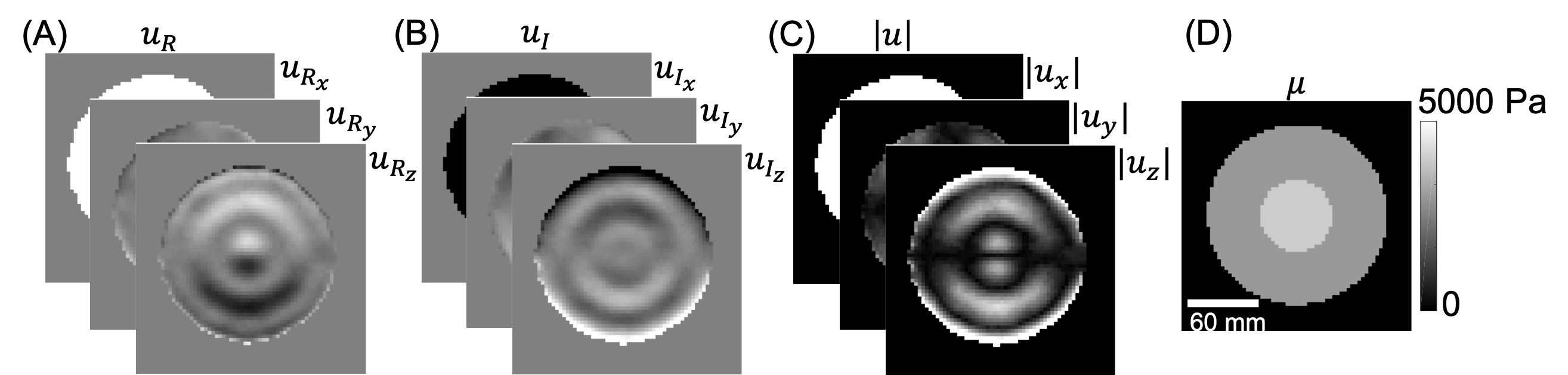}    
   	\caption{Sample entry from dataset. (A-B): Real and imaginary components of the complex valued displacement,$\protect\overrightarrow {u}$,  calculated from equations \ref{poroeq_a}-\ref{poroeq_b} given $\mu$ \ref{FIG:Motion}D. (C) Absolute displacement. (D) Ideal shear modulus image, $\mu$ }
\label{FIG:Motion}
\end{figure}

\section*{Convolution Neural Network}
As mentioned above, \citet{murphy2018artificial} demonstrated that a fully connected neural network can be used for mapping the 3D MRE displacement data into expected corresponding elastograms. Since fully-connected networks are not particularly suitable for two or more dimensional inputs, that work had to adopt a pre-processing step of partitioning input data into smaller 5x5 patches. This approach has the benefits of increasing the number of training examples while also reducing the number of connections between the input and hidden units and thus resulting in fewer network parameters. In fact, the approach is a hand-crafted means of implementing a convolution. So in this work, we opted to directly employ a convolutional layers in our network to circumvent the pre-processing steps. Additionally, we are able to make use of the recent advances in CNN implementations such as GPU computing. 

\begin{figure}[H]
\centering
  \includegraphics[width=0.95\textwidth]{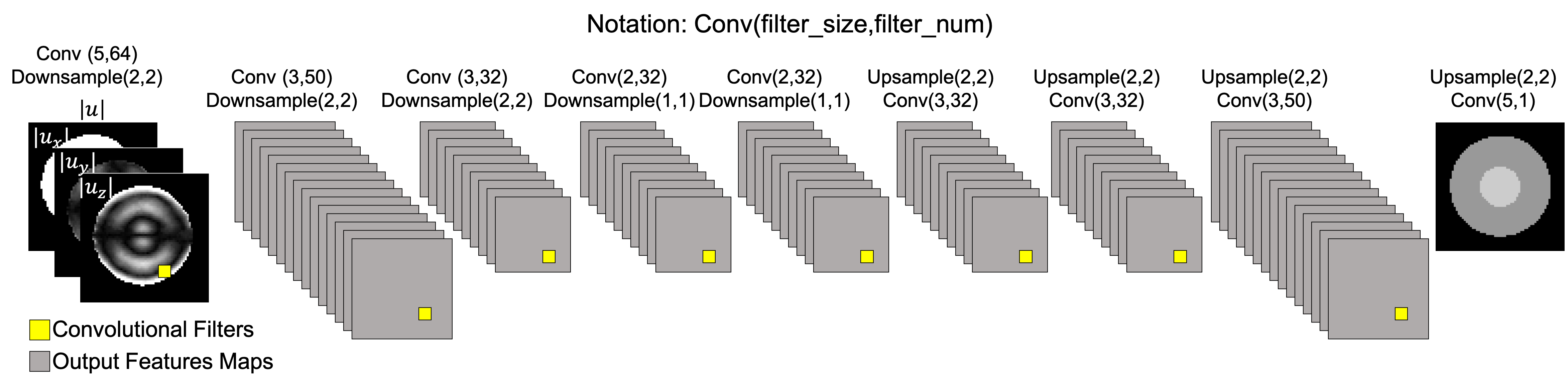}    
   	\caption{Convolutional neural network featuring an encoder-decoder architecture. 3D displacement data is encoded into an embedding, which is then decoded to reconstruct the 2D $\mu$ image. The network is trained using two loss functions: reconstruction loss and a custom MRE motivated loss. }
\label{FIG:Network}
\end{figure}

Our convolutional neural network features an encoder-decoder architecture. First, the encoder network takes in the 3D 64$\times$64$\times$3 displacement data and maps it down via 4 convolutions to an 8$\times$8 embedding. All downsampling is achieved through convolutions and no pooling layers are used in the encoder. The decoder then maps encoded embedding to the 2D 64$\times$64 elastogram. Fig. \ref{FIG:Network} illustrates this network architecture, detailing the number of units in the various layers.

Two loss functions are used in training out network. The primary loss function is the reconstruction loss which is also the mean squared error (MSE) loss. This ensures that elastogram images generated from the network match the target corresponding images from the nonlinear inversion method. Using MSE loss is typical for neural networks trained for image generation or reconstruction.

We introduce a secondary loss function which is motivated by established mechanical models of the MRE system. To derive this loss, equation \ref{poroeq_a} was simplified to the Navier's equation for viscoelastic material. This removed the last term $-(1-\beta)\nabla$ on the left hand side and replaced $p-\beta\rho_f$ on the right hand side with $\rho$. Assuming the absence of longitudinal waves in the displacement data, we further reduced the equation to a Helmholtz equation described by $\mu\nabla^2\overrightarrow{u}= -\omega^2\rho\overrightarrow{u}$. Thus, our secondary loss was given by: $|\mu\nabla^2\overrightarrow{u} - (-\omega^2\rho\overrightarrow{u})|_2$, and it ensured that our generated images were observed the underlying underlying mechanical relationship between $\overrightarrow{u}$ and $\mu$.

\section*{Network Training}
To train the proposed convolutional neural network, the dataset is split into two parts: 70\% for training, and 30\% for final testing. Furthermore, 30\% of the training set is also set aside as validation set for tuning training parameters such as batch size and number of epochs.

The network is implemented using Tensorflow and Keras. Training was performed using the adam optimizer, with a batch-size of 64 for 2,500 epochs. It took about 2 hours to train the network on a computer with an Nvidia Tesla K80 GPU, Intel Xeon E52640 2.6GHz CPU and 128GB RAM.

\section*{Results}
\begin{figure}[H]
\centering
  \includegraphics[width=1\textwidth]{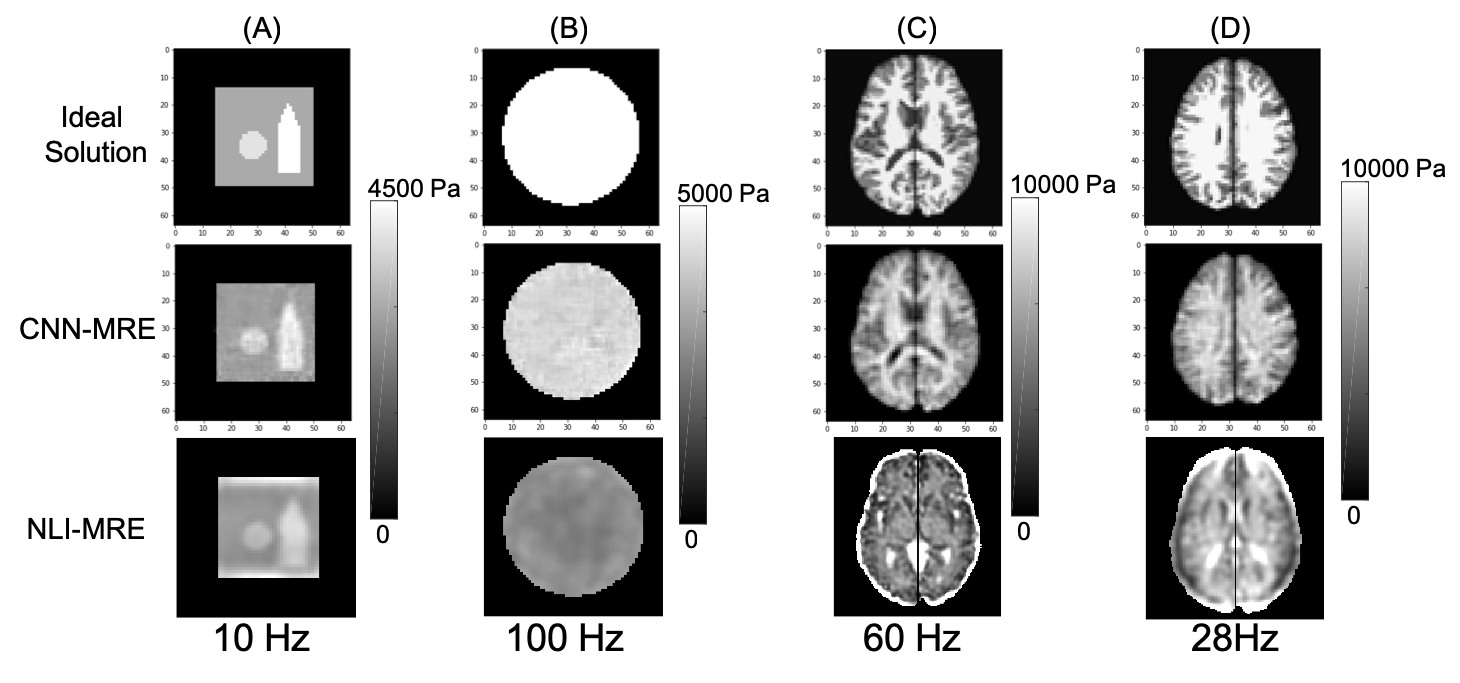}    
   	\caption{Reconstructed $\mu$ elastograms from the CNN network and a commonly used MRE inversion algorithm.}
\label{FIG:Results1}
\end{figure}

Figure \ref{FIG:Results1} shows the reconstructed images from the proposed CNN network for various actuation frequencies compared with estimated $\mu$ 
images from the standard non-linear inversion MRE algorithm. The neural network is able to recover no contrast images (\ref{FIG:Results1}B) as well the complexities and high stiffness variations like in the brain examples \ref{FIG:Results1}C and \ref{FIG:Results1}D. 

Supplemental figures show examples of reconstructions that deviated significantly from the expected solution. We noticed such deviations occur in low contrast entries where the network is unable to adequately distinguish between different regions.

\section*{Conclusion}
To our knowledge, \citet{kibria2018global} are the first to incorporate the use of CNN in the field of elastography to estimate displacements for strain imaging in ultrasound elastography. In this paper, we demonstrate the feasibility and the first application of CNN in MR elastography and its potential to provide clinically relevant mechanical properties of images. A major limitation so far have been the limited number and diversity of training examples. Future work will explore fine-tuning network hyper-parameters as well as incorporating training data from different inversion algorithms. Also, we would want to consider mechanical models with noise and study the influence of noise in the training and testing sets to enable more robust elastography neural networks. 

\subsubsection*{Acknowledgments}
We gratefully acknowledge support from NIH Grant No. R01-EB018230-01 and NIH Grant No. 1R21EB021456.

\bibliography{bibliography}

\section*{Supplemental Figures}

\begin{figure}[H]
    \includegraphics[width=0.48\textwidth]{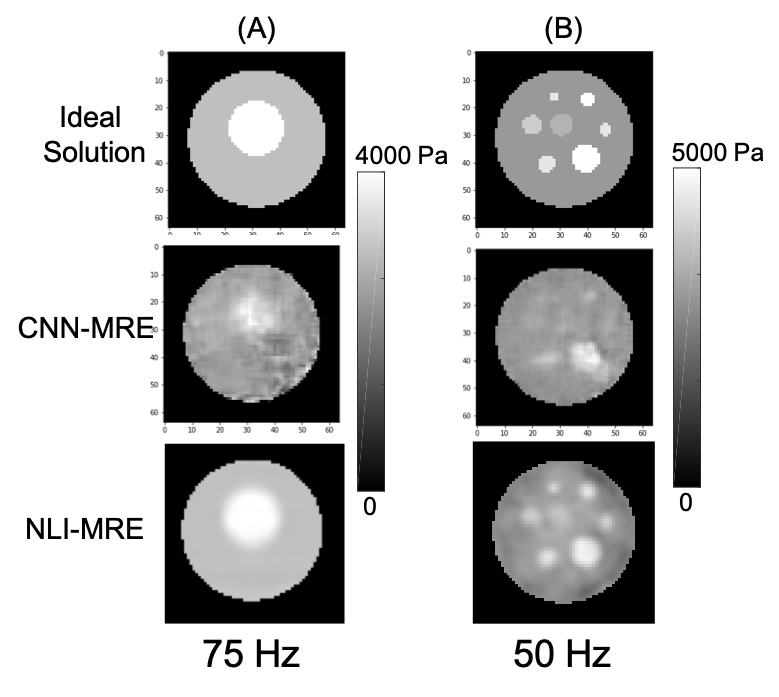}
    \caption{Reconstructed $\mu$ elastograms from the CNN network with poor stiffness resolution compared to the non-linear inversion estimates of $\mu$,}
  \label{FIG:Results2}
\end{figure}

\end{document}